\documentclass[12pt,preprint]{aastex}
\usepackage{amsmath, amsthm, amssymb}
\usepackage{ulem}
\newcommand{\rsun}{R_{\bigodot} }

\newcommand{\iec}{i.e., }

\newcommand{\of}{${F}_{\mathrm{open}}$} 

\begin{document}

\title{Surface flux transport modeling for solar cycles 15--21: effects of 
cycle-dependent tilt angles of sunspot groups}
\author{R.~H. Cameron}\email{cameron@mps.mpg.de}
\author{J. Jiang}
\author{D. Schmitt}
\and
\author{M. Sch\"ussler}

\affil{Max-Planck-Institut f\"ur Sonnensystemforschung, 
   37191 Katlenburg-Lindau, Germany}
\begin{abstract}
We model the surface magnetic field and open flux of the Sun from 1913 to 1986
using a surface flux transport model, which
includes the observed cycle-to-cycle variation of sunspot group tilts.
The model reproduces the empirically derived time evolution of the solar open magnetic flux,
and the reversal times of the polar fields. 
We find that both the polar field and the axial dipole moment resulting from this model 
around cycle minimum correlate with the strength of the following cycle.
 
\end{abstract}


\section{Introduction} 
The evolution of the large scale magnetic field  at the surface of the Sun can be modeled
using a two-dimensional surface flux transport model where the magnetic fields undergo a 
random walk due to supergranular flows 
\citep{Leighton64}, are advected by differential rotation and meridional circulation
\citep{Babcock61, Leighton64, Sheeley83}, and are
subject to a slow decay  from three-dimensional processes
\citep{Schrijver02, Baumann06}.  The surface field can be extrapolated out into the heliosphere, including the 
region near the Earth \citep{Wang00}. This makes the historic record of the magnetic 
environment of the Earth, as manifested in the geomagnetic perturbation indices,
 a valuable constraint on the variation of the magnetic fields of the Sun. 
The use of the surface flux transport model as the basis for field extrapolations has 
received some recent attention \citep{Mackay02, Wang02, Schuessler06, Jiang10}. 

The flux transport model has been refined over time: early work 
assumed time-independent flows and did not include the effects of radial diffusion.
It was found \citep{Lean02, Schrijver02} that the observed variation in 
the cycle amplitudes would  then lead to a secular drift of the 
polar field, in contradiction to observations. 
Two ways of extending the model were considered to cope with this problem: (1) making the meridional 
velocity time dependent \citep{Wang02_b}, and (2) assuming that the poloidal field decays 
with a timescale of about 5 years \citep{Schrijver02, Baumann06}.
Here we consider a third possibility, namely that the tilt angle of the groups, the subject of 
Joy's law, varies from cycle to cycle. 

Most surface flux transport studies have used fixed differential rotation and 
meridional flows. Two of the exceptions are the 
studies of \cite{Wang02_b} and \cite{Dikpati04} who considered cycle-to-cycle
changes of the meridional flow. \cite{Wang02_b}
 suggested that a suitably varying  time-dependent meridional flow 
would prevent a secular drift of the polar fields and thus allows them  to reverse every cycle despite  
large variations in the cycle amplitudes. 
 
Our study is motivated by the recent finding of \cite{Dasi-Espuig10} 
that the tilt angles of sunspot groups from the Mount Wilson Observatory 
and Kodaikanal  observations \citep{Howard84,Howard99, Sivaraman99}
show a cycle-to-cycle variation of  Joy's law. Further they showed that
the average tilt angle is negatively correlated with the 
strength of the cycle, \iec the tilt angle is smaller for stronger cycles.
A reduced tilt angle entails a smaller latitudinal separation between 
opposite polarity spots within a group, leading to reduced advection
and diffusion of following polarity magnetic flux towards the poles during strong cycles
\citep{Cameron07}.

Here we include the observed tilt angle variations 
as input to the flux transport model. Doing so requires us to 
reconsider the various parameters which go into the model (within
the range constrained by observations). In addition, we tentatively consider the 
effect of the observed inflows into active regions \citep{Haber04, Hindman04, Komm07}
which cause a reduced escape of flux from active regions \citep{De_Rosa06}.

The paper is organized as follows: Section 2 describes the flux transport
model, including a brief discussion of how well the model parameters are observationally known.
Section 3 outlines the observations to which the model's results are compared in
order to further constrain the parameters and test the model. In Section 4 we give 
the parameters for our reference case. A brief parameter study,
concentrating on the qualitative changes which occur as the different parameters
are varied, is presented in Section 5. Our conclusions are given in Section 6.

\section{Flux transport model}

The surface flux transport model describes the passive transport of 
the radial component of the magnetic field, $B$,  
on the solar surface under the effects of differential rotation, $\Omega$, 
meridional flow, $v$, and surface diffusivity, $\eta_{H}$, 
whilst gradually decaying owing to radial diffusion
\citep{Devore85,Sheeley85,Wang89,Mackay00,Schrijver02,Baumann04}. A 
source term, $S(\lambda,\phi,t)$, describes the 
emergence of new flux as a function of latitude, $\lambda$, and longitude, $\phi$. 
The governing equation is 

\begin{eqnarray}
\label{eqn:SFT}
\frac{\partial B}{\partial t}=& & -\Omega(\lambda) 
                       \frac{\partial B}{\partial \phi} 
         - \frac{1}{\rsun \cos\lambda}  
              \frac{\partial}{\partial \lambda}\left[v(\lambda) B \cos \lambda\right] \\ \nonumber
& & +\eta_{H} \left[\frac{1}{\rsun^2 \cos{\lambda}}
                \frac{\partial}{\partial \lambda}\left(\cos\lambda 
          \frac{\partial B}{\partial \lambda}\right) +
 \frac{1}{\rsun^2 \cos^2{\lambda}}\frac{\partial^2 B}{\partial \phi^2}\right]\\ \nonumber
& & +D(\eta_r)+ S(\lambda,\phi,t) ,
\end{eqnarray}
where $D$ is a linear operator describing the decay due to radial 
diffusion with radial diffusivity $\eta_r$. For $D$ we adopt the form used in \cite{Baumann06}. 
We use the time averaged (synodic) differential rotation profile given by  
\cite{Snodgrass83}: $\Omega(\lambda)=13.38-2.30 \sin^2 \lambda -1.62 \sin^4 \lambda$ (in $^{\circ}/$day). 
For the time-averaged meridional flow  we use the same profile as \cite{van_Ballegooijen98}, i.e.
\begin{equation}
v(\lambda)=
\begin{cases} 11 \sin(2.4 \lambda) \text{\,m\,s}^{-1} & \text{where} \vert \lambda \vert \le 75^{\circ}\\
              0 & \text{otherwise.}
\end{cases}
\label{eqn:mer}
\end{equation}
The two remaining parameters of the flux transport model are the horizontal 
and radial diffusivities, $\eta_{H}$ 
and $\eta_r$. The results of various attempts to measure $\eta_{H}$ from 
observation are summarized in Table 6.2 of \cite{Schrijver_book}. 
The values obtained from cross-correlation and object-tracking methods 
fall in the range $100-300$~km$^2$s$^{-1}$. 
We have used $\eta_{H}=250$~km$^{2}$s$^{-1}$ 
for our reference value in Section 4; this value lies within the range of the 
observations, but we also consider the effect of varying it in Section 5.

Much less is known about $\eta_r$. This term was introduced by \cite{Baumann06} 
 to account for the 3D radial diffusion of the magnetic field and to 
obtain regularly reversing polar fields for cycles of varying amplitude 
in the absence of variations of the meridional flow. Its physical motivation 
is that the Sun's magnetic field is three-dimensional and thus has more modes of decay
than are captured by the two-dimensional surface diffusion.  
We find here that the results with $\eta_r=0$ match the 
observations well (including having the polar fields reverse each cycle)
when we include the observed tilt angle variations. We thus take 
$\eta_r=0$ as our reference value and consider other values in 
Section 5.

For the source term $S(\lambda,\phi,t)$ in Equation~\ref{eqn:SFT}  
we follow \cite{van_Ballegooijen98} and \cite{Baumann04} and consider 
new flux to emerge in the form of  of opposite polarity patches. The positive-polarity patch 
is centered on latitude $\lambda_+$ and longitude $\phi_+$, the negative patch 
at  $(\lambda_-,\phi_-)$. The field of each new bipole is given by 
$B=B^+-B^-$ with
\begin{eqnarray}
B^{\pm}(\lambda,\phi)=B_{\mathrm{max}} \left(\frac{0.4 \Delta\beta}{\delta}\right)^2
          \mathrm{exp}({2[1-\cos(\beta_{\pm}(\lambda,\phi))/\delta^2]}),
\end{eqnarray}
where  $\beta_{\pm}(\lambda,\phi)$ are the heliocentric angles between $(\lambda,\phi)$
and $(\lambda_{\pm},\phi_{\pm})$, respectively and 
$\Delta \beta=\beta_+(\lambda_-,\phi_-)$ is the separation between the two polarities,
and $\delta=4^\circ$ is the size of the individual polarity patches.
For the purposes of comparing the flux transport simulations with observations it is 
necessary to  connect $S$ closely  to the actual observations.
We use sunspot group areas and locations corresponding to their time of maximum area  
from \url{http://solarscience.msfc.nasa.gov/greenwch.shtml} (based on the Greenwich 
photoheliographic maps from 1874 to 1976 and USAF/NOAA SOON data thereafter)
as proxies for emerging flux.

\begin{figure}
\epsscale{0.95}
\plotone{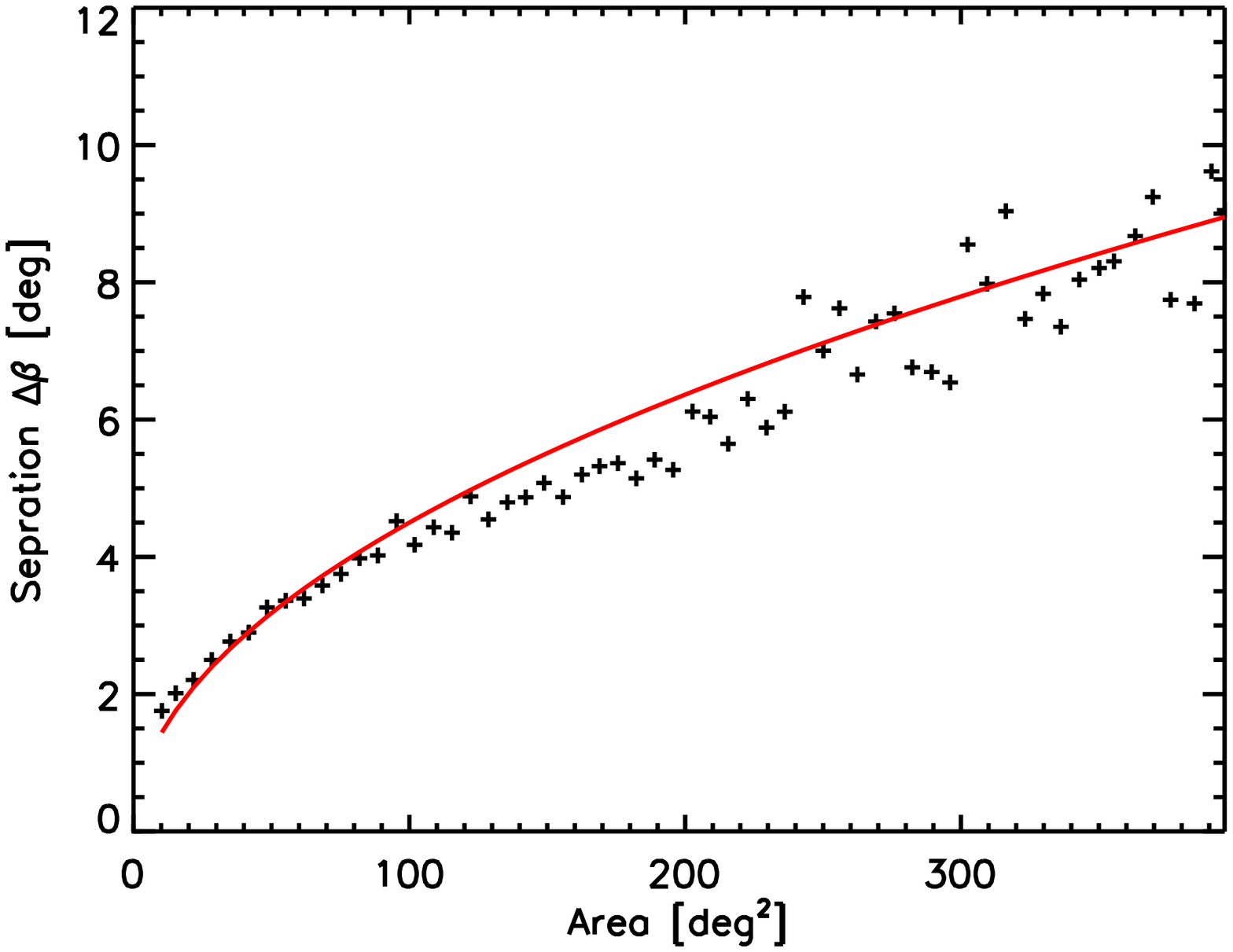}
\plotone{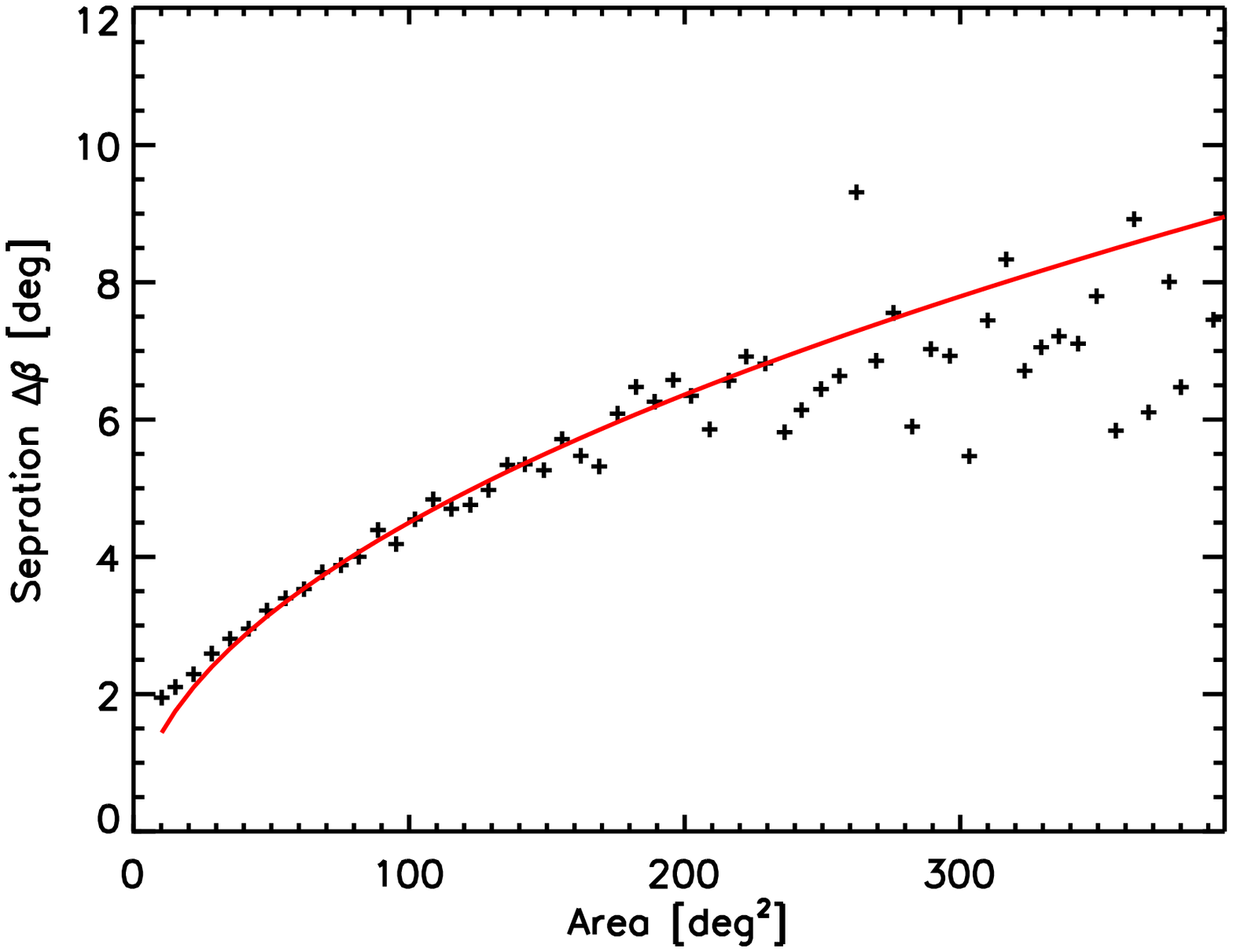}
\caption{The separation between the following and leading polarities as a function of sunspot
group area. The values are from the Kodaikanal (upper panel) 
and  Mount Wilson Observatory data sets (lower panel). 
The solid red curve is the fit $\Delta\beta=0.45 A_R^{1/2}$.}
\label{fig:calib}
\end{figure}

The  Greenwich/USAF/NOAA record contains the locations and areas of sunspots groups,
but no magnetic polarity information. 
We use the location and areas to construct bipolar magnetic regions with the form described
by Equation 3.
The location of the bipoles, $(\lambda_{\pm},\phi_{\pm})$, in the 
northern hemisphere are given by 
\begin{eqnarray}
\lambda_{\pm}&=&\lambda_m\pm (-1)^n 0.5 \Delta\beta \sin\alpha\\
\phi_{\pm}&=&      \phi_m\mp (-1)^n 0.5 \Delta\beta \cos\alpha (\cos \lambda)^{-1},
\end{eqnarray} 
and those in the southern hemisphere by
\begin{eqnarray}
\lambda_{\pm}&=&\lambda_m\pm (-1)^n 0.5 \Delta\beta \sin\alpha\\
\phi_{\pm}&=&      \phi_m\pm (-1)^n 0.5 \Delta\beta \cos\alpha (\cos \lambda)^{-1}.
\end{eqnarray} 
Here $(\lambda_m,\phi_m)$ is the central location of the group from the Greenwich/USAF/NOAA record,
$\alpha$ the tilt angle with respect to the azimuthal direction,
and $n$ the cycle number. 
The separation, $\Delta \beta$, 
between the two polarities is taken to be $\Delta\beta=0.45 A_R^{1/2}$ where $A_R$
is the total area of the active region. We estimate the total flux of an  active region by considering 
its total area, $A_R$, to be the sum of the area covered by the sunspots $A_s$ and the plage $A_p$ 
using the observed relationship

\begin{equation}
A_R=A_s+A_p=A_s+414+21 A_s-0.0036A_s^2,
\end{equation} 
\citep{Chapman97},
where all areas are measured in millionths of a solar hemisphere.  
The coefficient 0.45 was determined by using the sunspot group data
from  Mount Wilson (covering the period from 1917 to 1985)  and 
Kodaikanal (covering the period from 1906 to 1987). 
The data sets are described in \cite{Howard84} and \cite{Sivaraman99}
and are available from http://ngdc.noaa.gov/stp/SOLAR/ftpsunspotregions.html.
Both data sets include umbral areas $A_U$ and separations $\Delta\beta$ between the 
``centers of mass'' of the leading and following spots. We converted the umbral areas to sunspot areas
using the results of \cite{Brandt90}, and  from there to $A_R$ using Equation 8. Figure~\ref{fig:calib}
shows the average (over 7-degree bins) of $\Delta \beta$ from each data set together
with the fit curve.

\begin{figure}
\epsscale{0.75}
\plotone{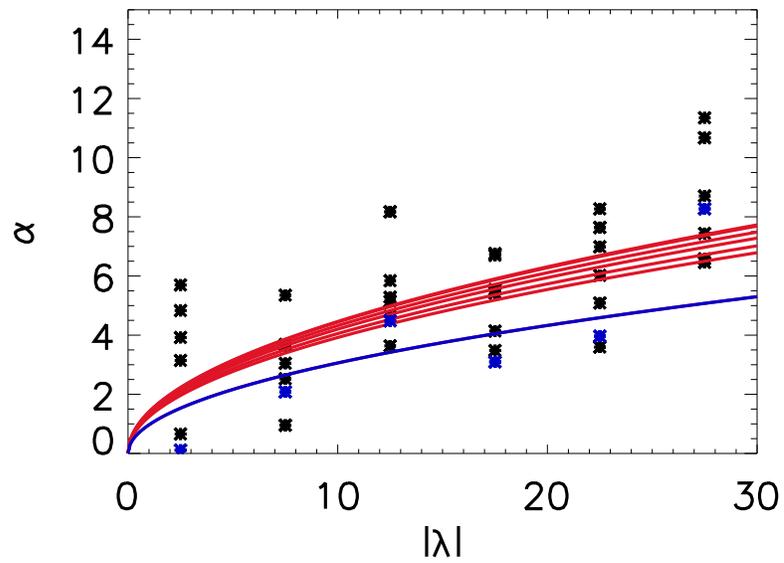}
\caption{Area weighted tilt angle $(\sum_i A_i \alpha_i)/(\sum_i A_i)$ as a function of 
sunspot group latitude for cycles 15 through 21 on the basis of the
combined Mt. Wilson and Kodaikanal datasets. The asterisks represent the average values for 5 
degree bins for each cycle. The curves represent fits of the form 
$\alpha=T_n \sqrt{\vert \lambda \vert}$ for each cycle $n$. The blue asterisks and curve indicate cycle 19.}
\label{fig_data2}
\end{figure}

\begin{figure}
\epsscale{0.75}
\plotone{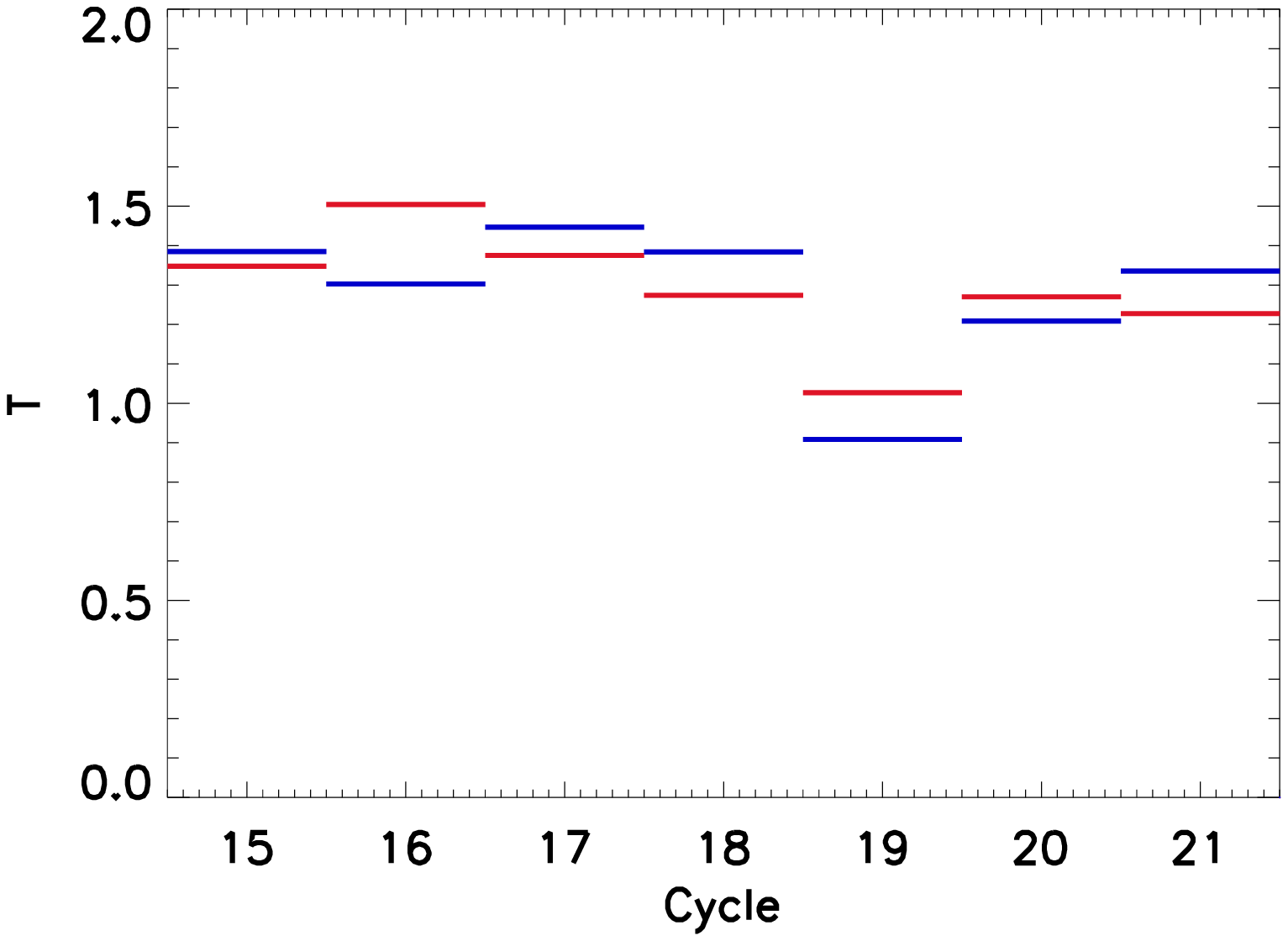}
\caption{The cycle dependence of the factor $T_n$ in the fit curves of Figure~\ref{fig_data2}.
Shown are the values derived from the Mt. Wilson data (blue) the Kodaikanal
data (red).}
\label{fig_data3}
\end{figure}

The next step is to specify the tilt angle, $\alpha$, including its 
cycle-to-cycle variations. As noted in \cite{Baumann04} the
polar fields are essentially proportional to $\alpha$, so that these variations might strongly 
affect the results. 
We use the tilt angles provided by the sunspot group data from Mount Wilson and from Kodaikanal. 
Since these observations cover only part of the sunspot groups in the combined 
Greenwich/USA/NOAA dataset that we use for the source, we take cycle-averaged properties for the
tilt angle as a function of latitude.
The asterisks in Figure~\ref{fig_data2} show the binned cycle averages of the tilt angle weighted with the 
group areas. We fit the data from each cycle to the form   
$\alpha=T_n \sqrt{\vert \lambda \vert}$, 
where $n$ is the cycle number. 
We calculate $T_n$ by
\begin{equation}
T_n=\frac{\sum_i A_i \alpha_i}{\sum_i A_i \sqrt{\vert \lambda_i \vert }}
\end{equation}
where the summation is over all spot groups in cycle $n$; $A_i$ is area of the $i$-th spot
group and $\lambda_i$ is its latitude. 
If the tilt angle of each group of cycle $n$ is written as 
$\alpha_i=T_n \sqrt{\vert \lambda_i \vert} +\epsilon_i$, then the above estimate for $T_n$ implies
$\sum_i A_i \epsilon_i=0$ \iec the area-weighted sum of the deviations from the fit curve is zero.
We see in  Figure~\ref{fig_data2} that the square-root profile produces a reasonable
fit. Furthermore cycle 19 is systematically low across a broad range of latitudes.
Figure \ref{fig_data3} shows $T_n$ for both the Mount Wilson and Kodaikanal data sets. 

Observed localized inflows associated with active regions 
\citep{Haber04, Hindman04, Komm07} are important for the evolution of the surface magnetic fields \citep{De_Rosa06}
and should be included in the model. The effect of the inflows is to reduce the rate 
of expansion of the active region flux. This reduces the latitudinal separation of the polarities and thus  
the amount of net flux which can migrate to the poles. Additionally the large-scale inflows 
into the activity belt \citep{Gizon08} affects the polar fields.
\cite{Jiang10b} have quantitatively studied the  effects of such
inflows on the polar fields, finding that a 5 m/s global inflow such as was observed during
cycle 23 reduces the polar fields by 18\%. The flows into individual active regions are 
stronger than this global-scale inflow and thus have a bigger effect \citep{De_Rosa06}.
In this paper we have tentatively 
introduced a time-independent parameter, $g$, which we use to scale the observed tilt angles
of the sunspot groups. The main effect of $g$ is to scale the amount of 
flux which reaches the pole against the amount of flux which emerges in the sunspot groups.
We comment that the introduction of this parameter also deals with the uncertainty of how the 
observed group tilt angles (which are based upon the position of sunspots in continuum intensity images)
are related to the tilt angles between the opposite polarity patches of the active region.

Finally we determine $B_{\mathrm{max}}$ in Equation 3 by matching the unsigned total observed flux 
from the Mount Wilson and Wilcox observatories with the simulation results.

For the initial magnetic field distribution at the start of the simulation  we follow \cite{van_Ballegooijen98}
and use an axisymmetric solution to Equation 1 with $B=\pm B_0$ at the poles, and which evolves almost entirely on the 
slow diffusive timescale: 
\begin{equation}
B=\begin{cases}\mathrm{sign}(\lambda) B_0 \exp{\left(\frac{-11 \mathrm{ms}^{-1} \times \rsun}{2.4 \eta_{H}} 
                     \left(1+\cos 2.4 \lambda \right)\right)} & \text{if } \vert \lambda \vert < 75^{\circ} \\
               \mathrm{sign}(\lambda) B_0&\text{otherwise}.
\end{cases}
\end{equation}
In the absence of sources, the evolution of this initial condition is  
dominated by the slow decay of the global field (with an  e-folding time of approximately 4000 years).
This choice for the functional form of the initial condition is arbitrary. The time between flux emerging and its reaching the poles 
is on the order of 5 years, so the polar fields for the first 5 years or so are
strongly affected by the chosen form for the initial condition. 
We therefore exclude the first polar field maximum from our analysis of the results.

\section{Open flux  and timing of the polar reversals}
The model described above requires emerging bipolar magnetic region areas, locations, and tilt 
angles as input. We have these data for the period between 1913 and 1986.
The flux transport model gives as output the radial component of the magnetic field on the 
solar surface. Throughout most of the period covered by the simulations, observational magnetogram data are 
unavailable for comparing against the results of the simulations. We therefore consider
the Sun's open flux, \of, which has been inferred from the $aa$-index of geomagnetic variations (and its extensions) 
from 1842 onwards \citep{Lockwood99, Lockwood03}. 
To obtain \of\ from the simulation, we take the surface 
distribution of $B$ and use the current sheet source surface model \citep{Zhao95b, Zhao95, Zhao02} 
to extrapolate the solar surface field out into the heliosphere  \citep[see][]{Schuessler06, Jiang10}. 
The results of the extrapolation are dependent on the assumed value of the 
`cusp radius', $R_{\mathrm{cusp}}$, 
which is the radial distance beyond which all the field lines are open.  

We also compare the simulation results against the timing of the polar field reversals, which have been inferred by 
\cite{Makarov03} from polar filament observations from 1870 to 2001.

\section{Reference case}
\begin{figure}
\epsscale{0.75}
\plotone{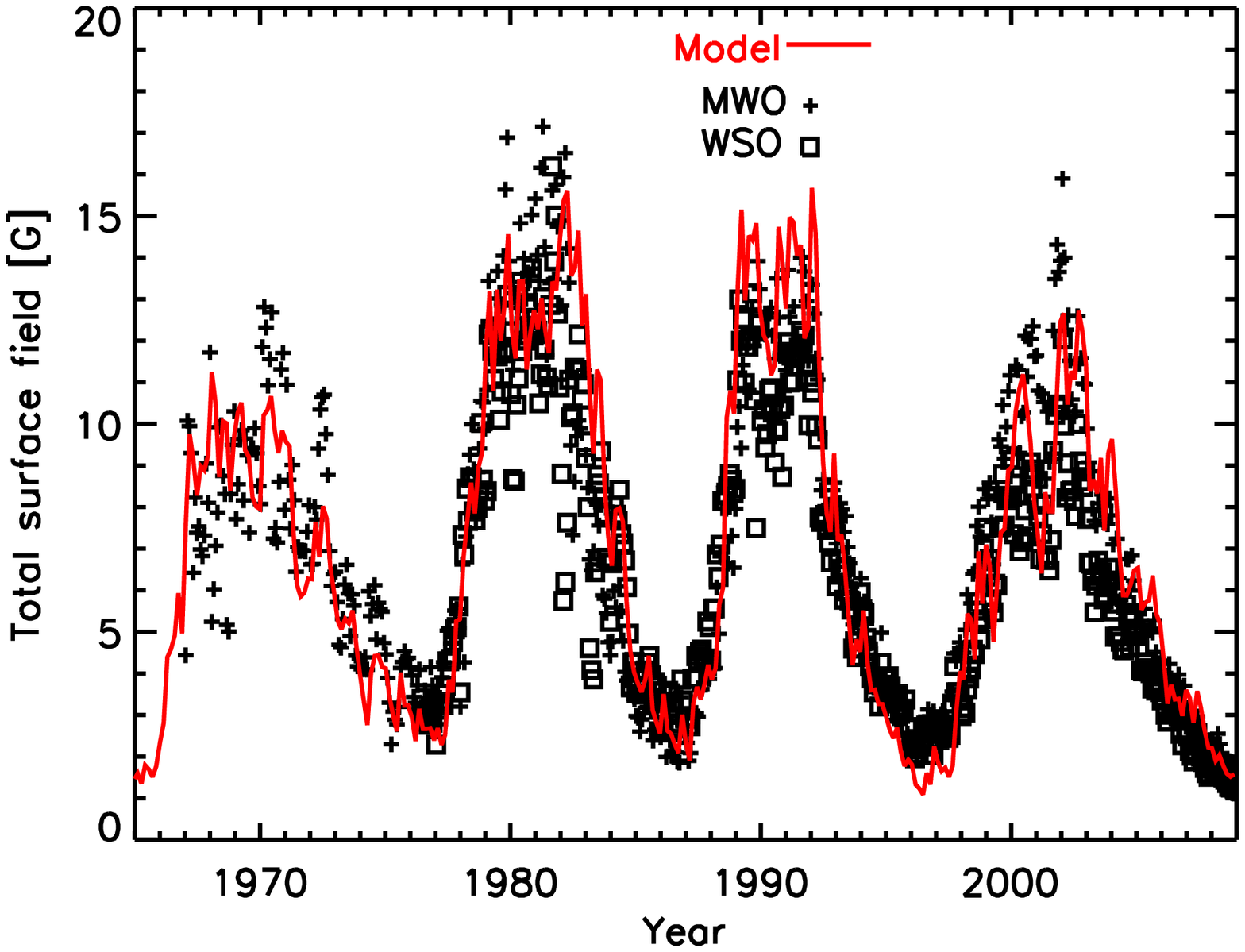}
\caption{Average unsigned magnetic field from the simulation with $B_{\mathrm{max}}=374$~G 
(solid curve) compared with observations from the Mount Wilson (plus signs) and 
Wilcox Solar Observatories (squares).}
\label{fig:calib_Bmax}
\end{figure}

\begin{figure}
\epsscale{0.75}
\plotone{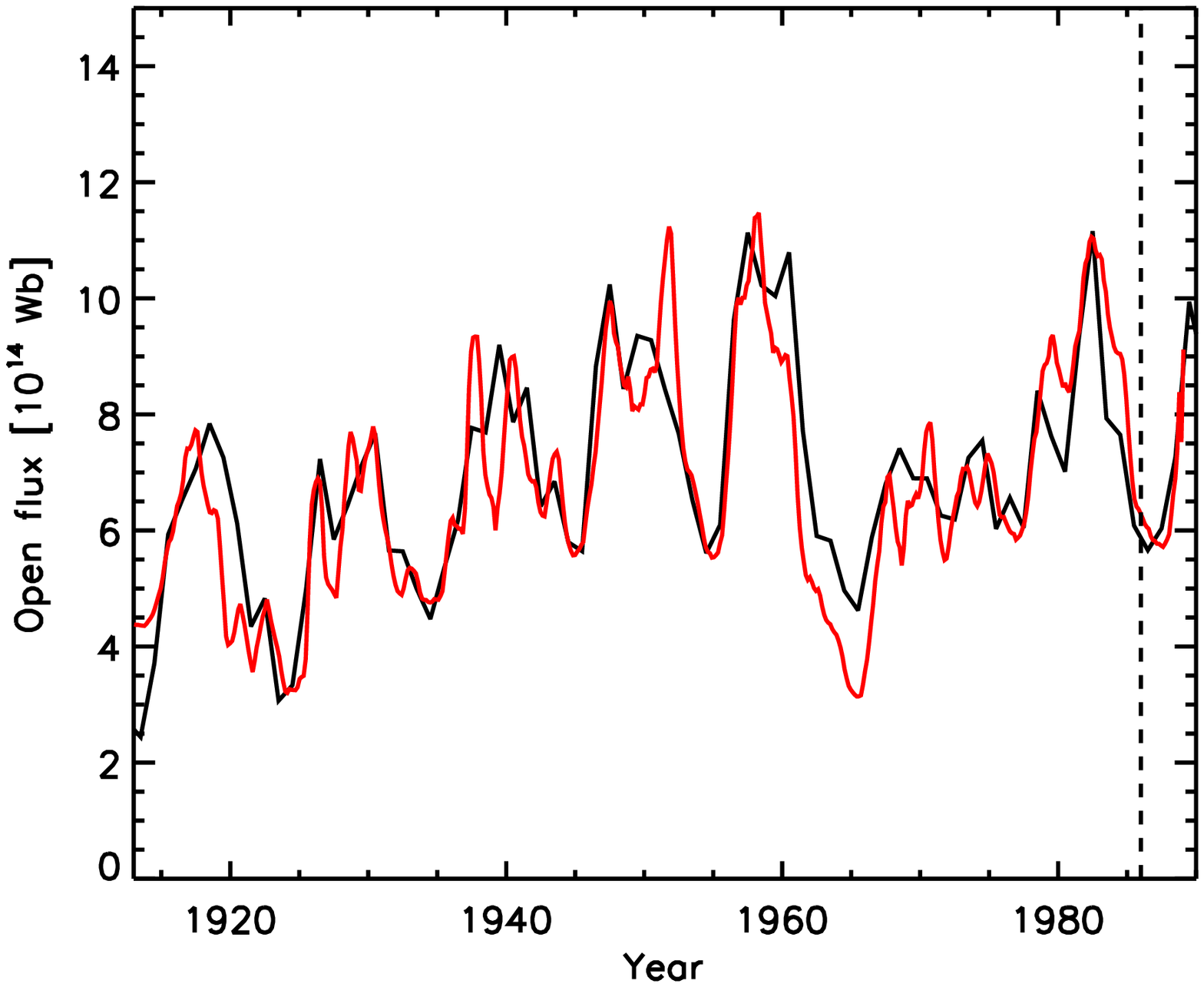}
\caption{Open flux inferred from the $aa$-index \citep[][black]{Lockwood03} and result
from the surface flux transport model and extrapolation (red) for the reference case.
The dashed vertical black line refers to  the time beyond which we have no observed values 
of the tilt angle.}
\label{fig:ref}
\end{figure}

\begin{figure}
\epsscale{0.75}
\plotone{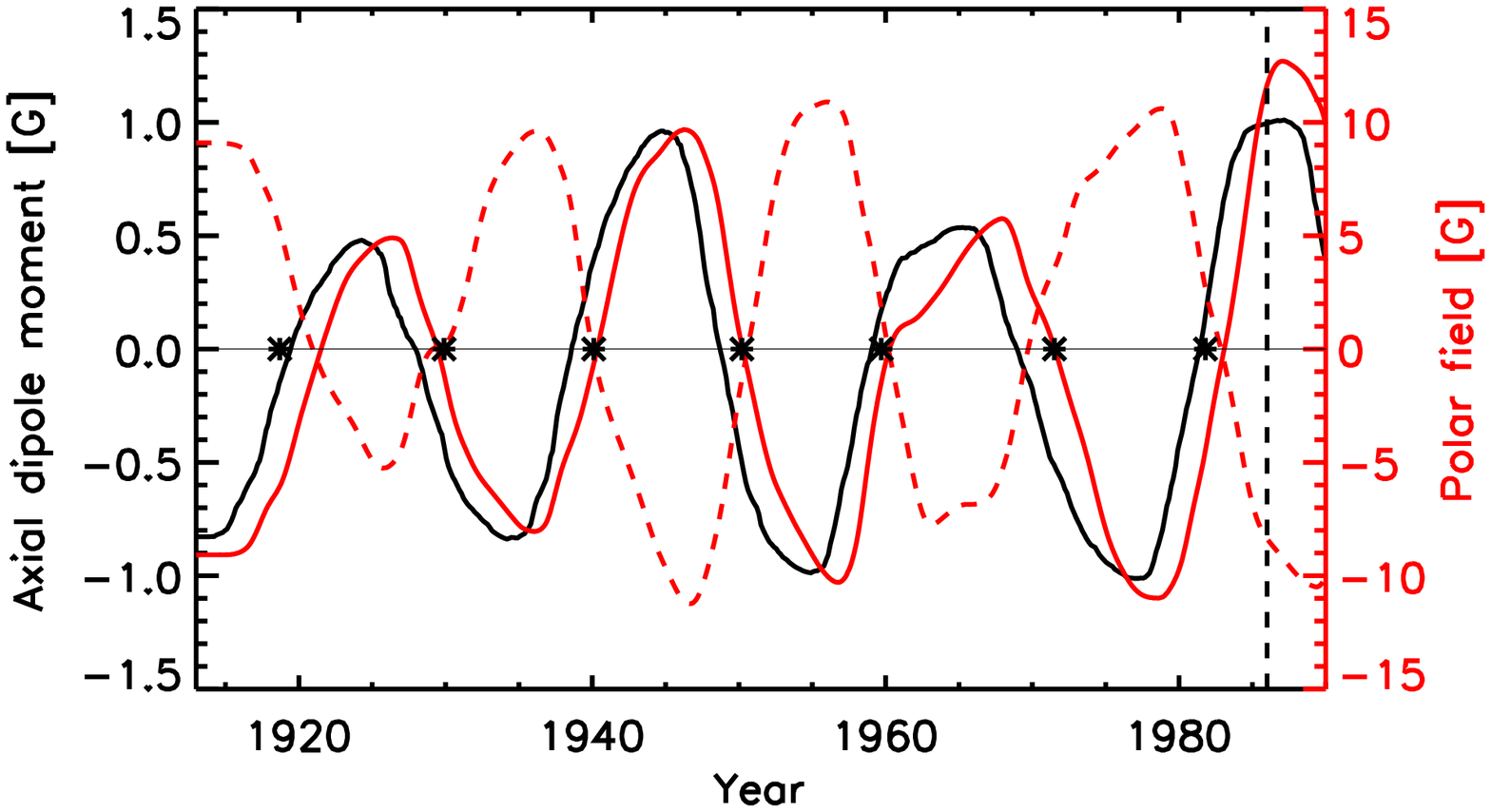}
\plotone{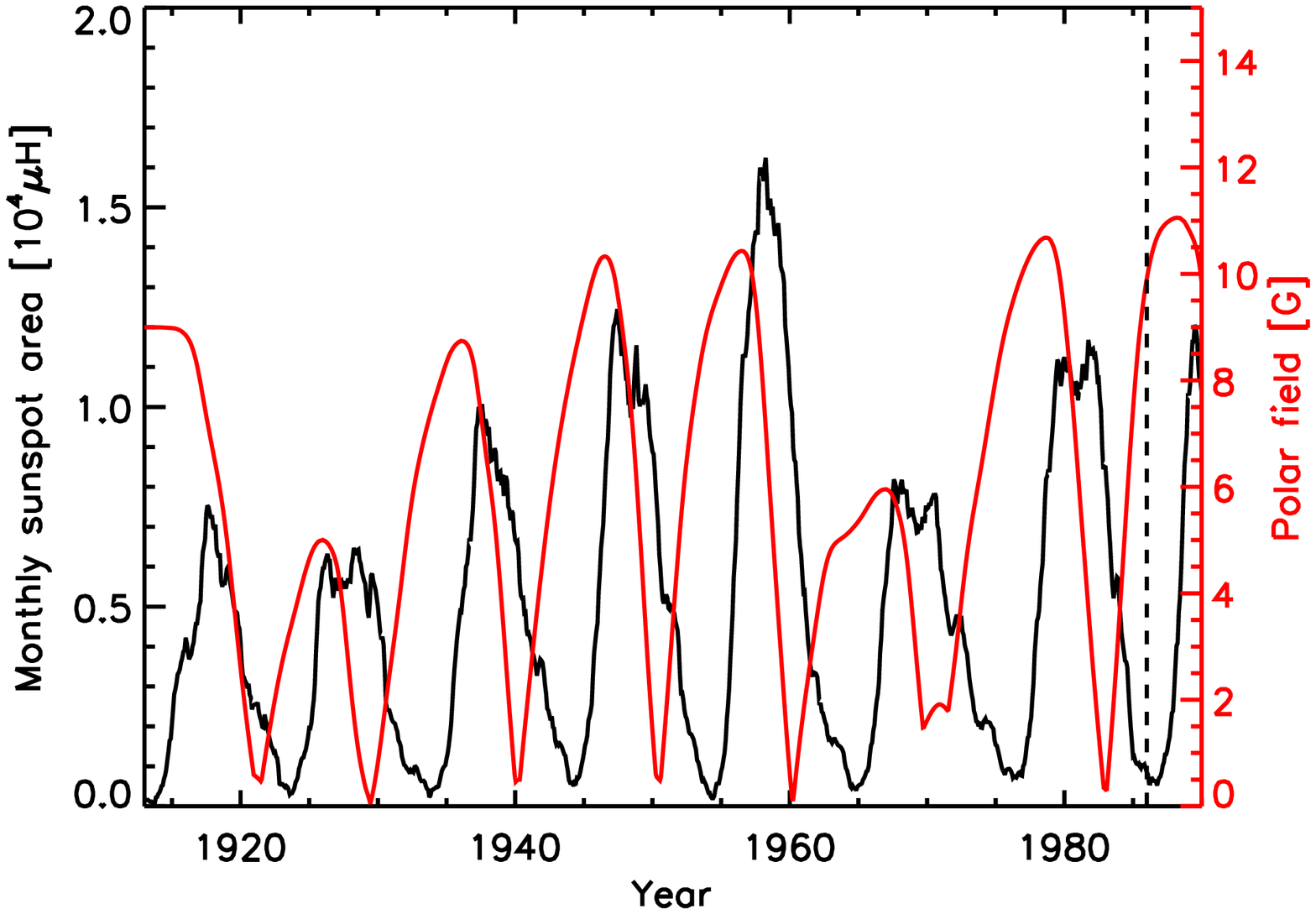}
\caption{Upper panel: axial dipole moment (black) together with the north (solid red) and  
south (dashed red) polar field strengths from the flux transport simulation. 
The black asterisks indicate the inferred times of polar field reversals
from \cite{Makarov03}.
The dashed vertical black line refers to  the time beyond which we have no observed values 
of the tilt angle.
Lower panel: average of the unsigned polar field strength from the flux transport model 
(red) and observed sunspot area (black).
}
\label{fig:ref_pol}
\end{figure}

We here give the results for the parameter set 
$\eta_{H}=250$ km$^2$s$^{-1}$, $g=0.7$, $B_0=-10.2$ G, $\eta_r=0$ and $R_{\mathrm{cusp}}=1.55 \rsun$. 
The value $B_{\mathrm{max}}=374$~G was found by matching to the total observed unsigned magnetic flux
from the Mount Wilson and Wilcox Solar Observatories (see Figure~\ref{fig:calib_Bmax}). 
With these parameters, the model reproduces well the open flux \of~ inferred by \cite{Lockwood03} 
as shown in Figure~\ref{fig:ref}. This applies to the phases as well as the 
amplitudes of both the maxima and minima of the inferred open flux. 
The corresponding evolution of the  polar field (defined as the average field above $\pm 75^{\circ}$ latitude) 
and axial dipole moment are shown in the upper panel of Figure~\ref{fig:ref_pol}. 
The polar field closely follows the axial 
dipole moment, with a delay of several years. This is understandable as the dipole 
moment reacts more quickly to flux transport across the equator, which then takes several 
additional years to reach the polar latitudes ($>75^{\circ}$). 
The simulated polar fields reverse for all cycles. Without the
cycle-dependent variations of the tilt angle the weak cycle 20 would have been unable
to offset the polar field after cycle 19. 
It was this type of problem which led \cite{Schrijver02} and \cite{Baumann06} to introduce
a decay term. Here we achieve a good agreement with the observations even
without such a term because of the anti-correlation in the
observed tilt angles and cycle strengths \citep{Dasi-Espuig10}.
The asterisks in the upper panel of Figure 6 indicate the timings of the polar
reversals as derived by \cite{Makarov03} from  H$\alpha$ polar filament data. 
The reversal times are reasonably 
well reproduced, except for the first reversal which is still affected by 
arbitrary form of the initial condition.

\begin{figure}
\epsscale{0.5}
\plotone{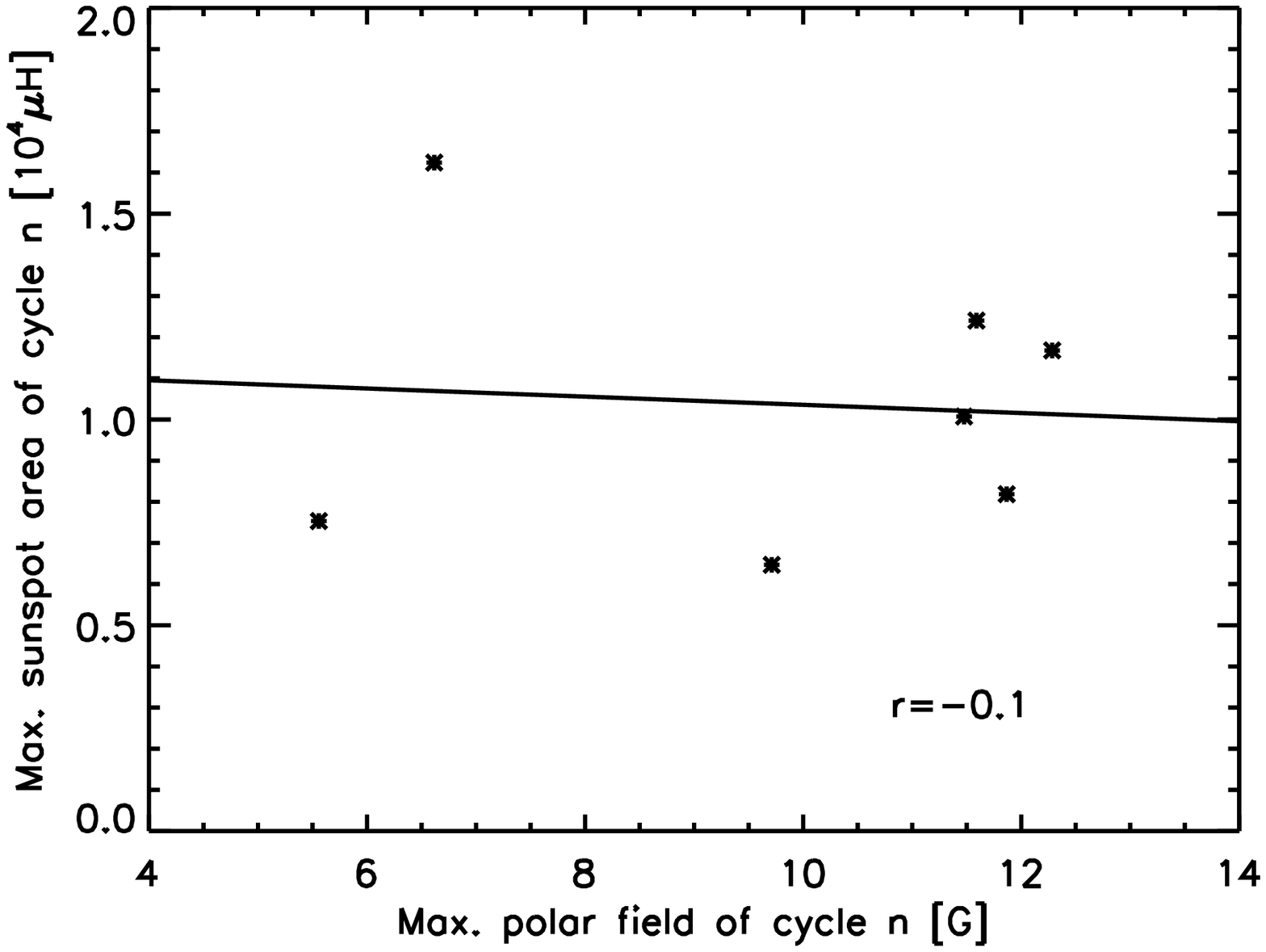}
\plotone{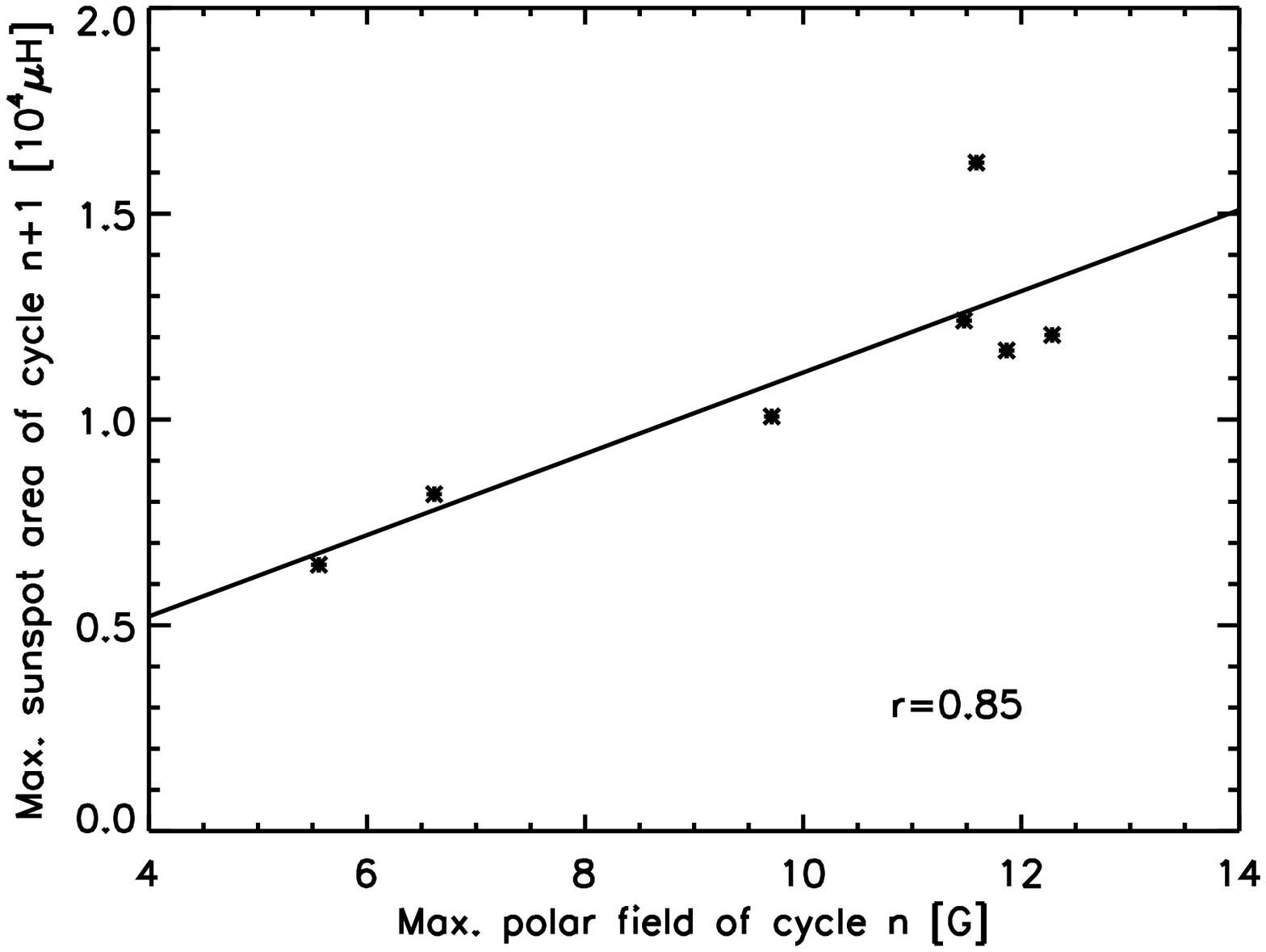}
\caption{
Correlation between the observed maximum  polar field from the flux transport model
and and the preceding (upper panel) and subsequent (lower panel) cycle's maximum sunspot area. 
The asterisks represent the values for individual cycles: the solid line is a linear fit.
The maximum sunspot area is uncorrelated ($r=-0.1$) to the  subsequent polar field maximum,
but strongly correlated ($r=0.85$) to the preceding polar field maximum. 
The first maximum of the polar field has been omitted from the analysis since it is affected by the
arbitrariness of the initial field.}
\label{fig:corr}
\end{figure}

The maximum of the dipole moment during the activity minimum
between cycles 19 and 20 is much lower than that between cycles 18 and 19.
This is primarily because the average tilt angle of cycle 19 was substantially lower than 
that of cycle 18 (see Figures 2 and 3).
On the other hand, the dipole moment between cycles 20 and 21 is again high,
because the average the tilt angles of cycle 20 was high. 
The lower panel of Figure~\ref{fig:ref_pol} presents the time evolution of the 
unsigned north-south averaged polar field from the simulation as well as the observed
strength of the solar activity. Ignoring the first cycle, which is strongly
affected by the initial condition, the polar field can be seen to anticipate
the peaks of solar activity. We have quantified this relationship by calculating the
correlation between the strength of peaks the (unsigned north-south averaged) 
polar field and the strength of the adjacent activity maxima. 
As can be seen in Figure~\ref{fig:corr}, the maxima of the polar field between 
cycles $n$ and $n+1$  reflect the amplitude of cycle $n+1$ much better than cycle $n$. 
The correlation coefficient for the relationship between the polar field
and the amplitude of the next cycle is 0.85. 
This is in contrast to the result of  \cite{Cameron07} who did not consider a 
cycle-to-cycle dependent tilt angle; they found that 
the polar fields and strength of the polar field closely follow the previous cycle.
This indicates the importance of the cycle-to-cycle variations in the tilt angle.

Some observational evidence concerning the correlation between the polar field and the strength 
of the previous and subsequent cycles has been previously considered, e.g, by 
\cite{Schatten78,Layden91,Svalgaard05,Jiang07}. 
The existence of a correlation between the polar field and the strength of the
next cycle is evidence in favour of a Bacock-Leighton-type dynamo.
Within the context of such dynamos, the correlation constrains the subsurface 
dynamics \citep[see, for example,][]{Yeates08}.

\section{Parameter dependence}
\begin{figure}
\epsscale{1}
\plottwo{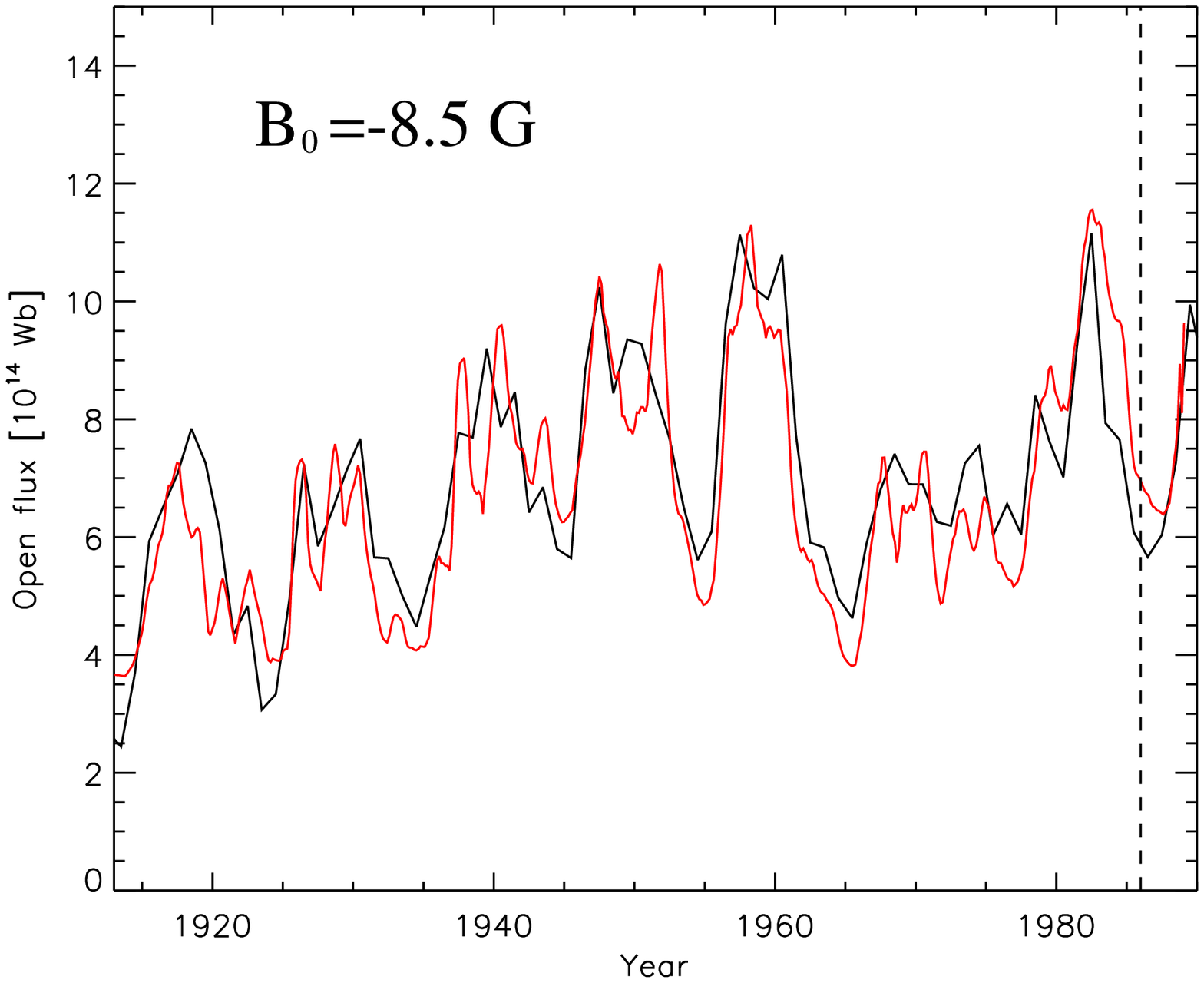}{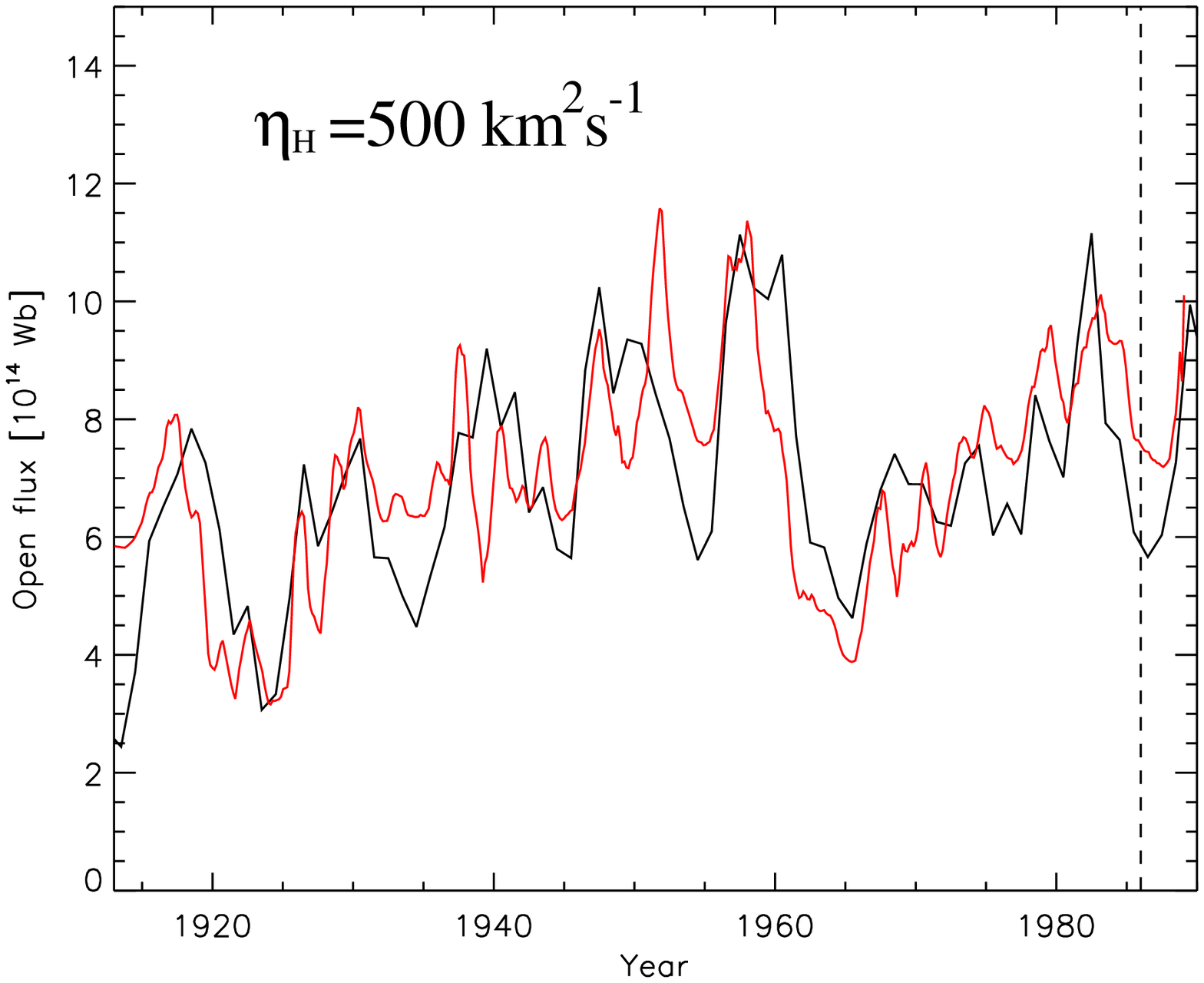}

\plottwo{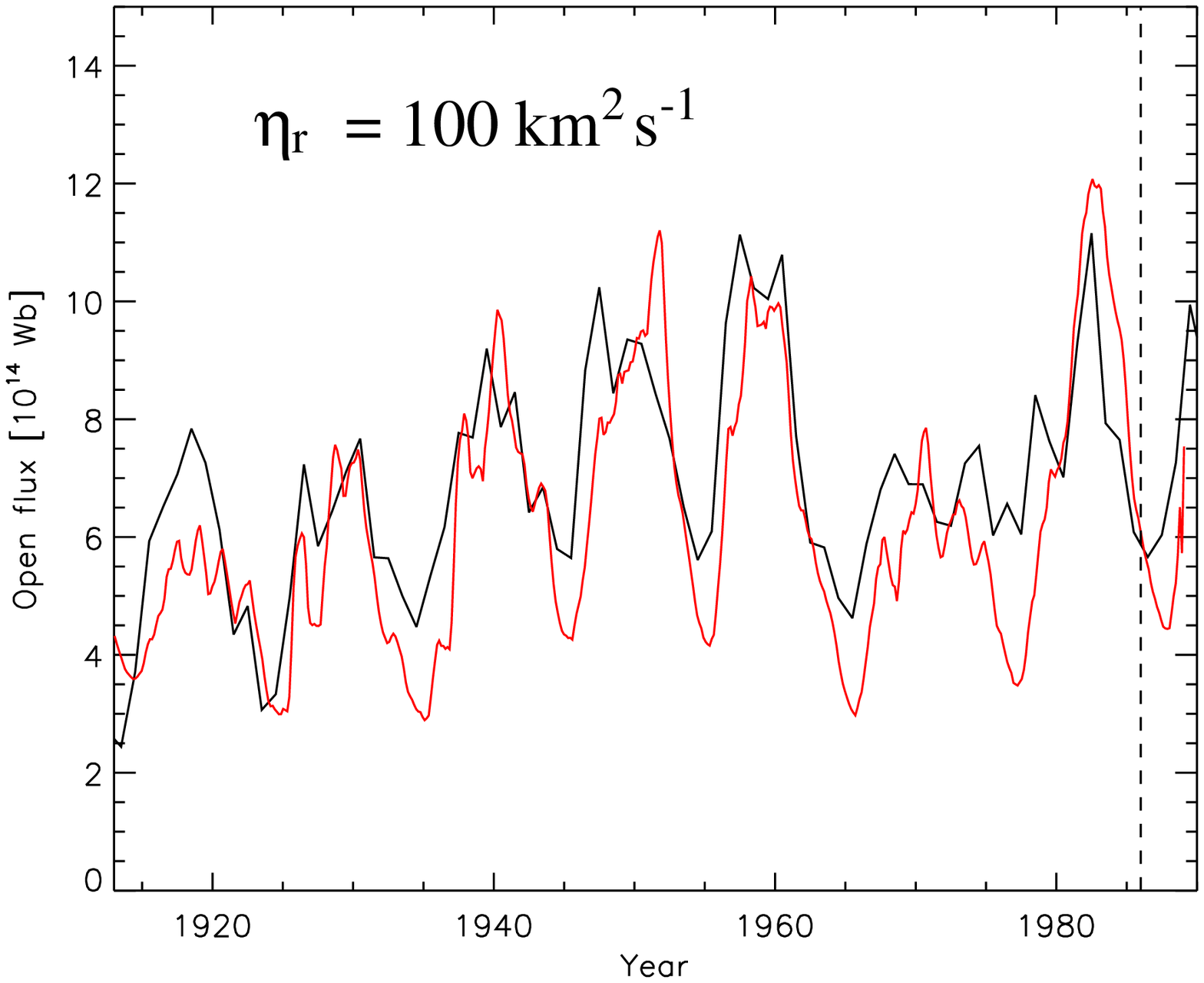}{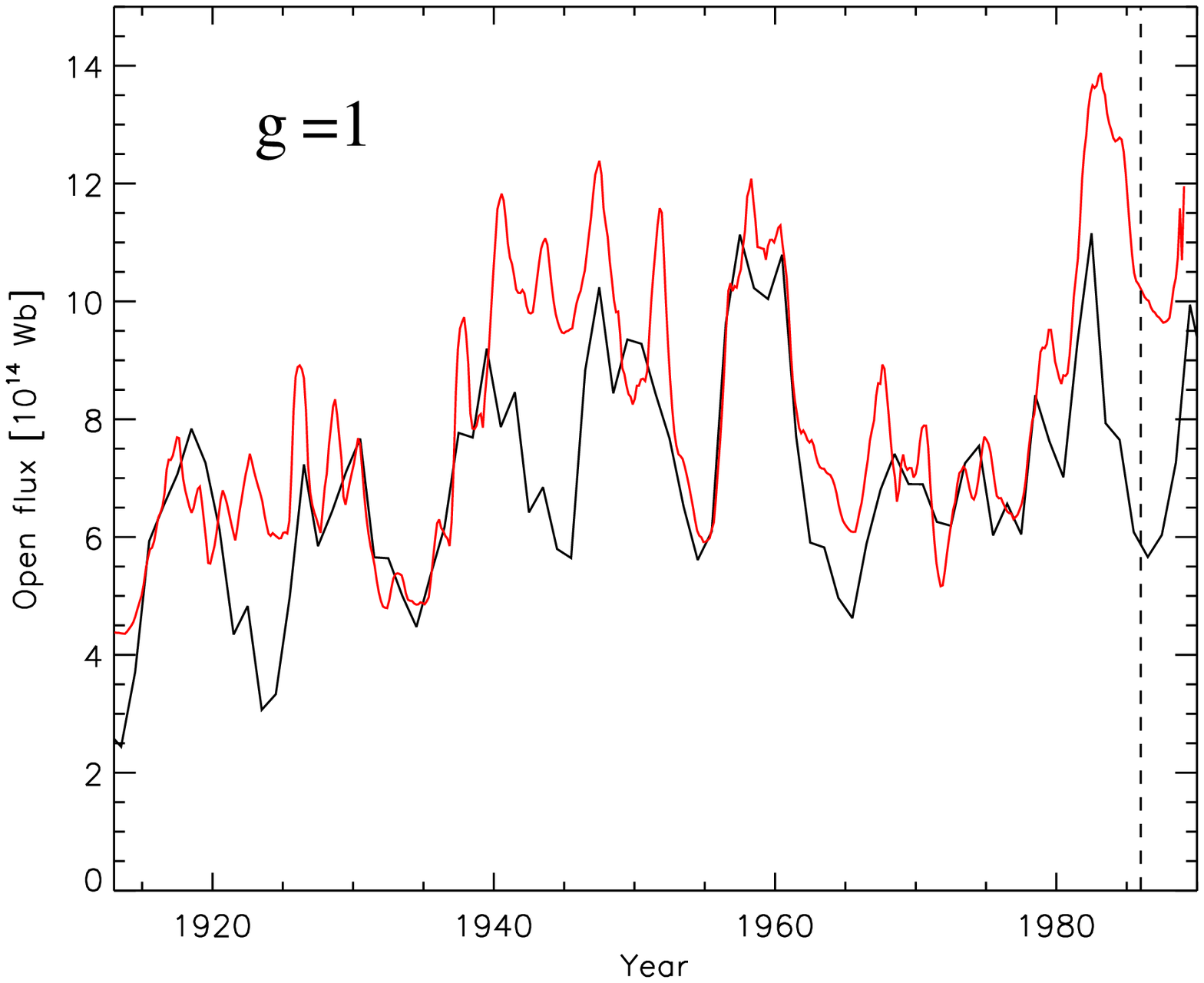}

\plottwo{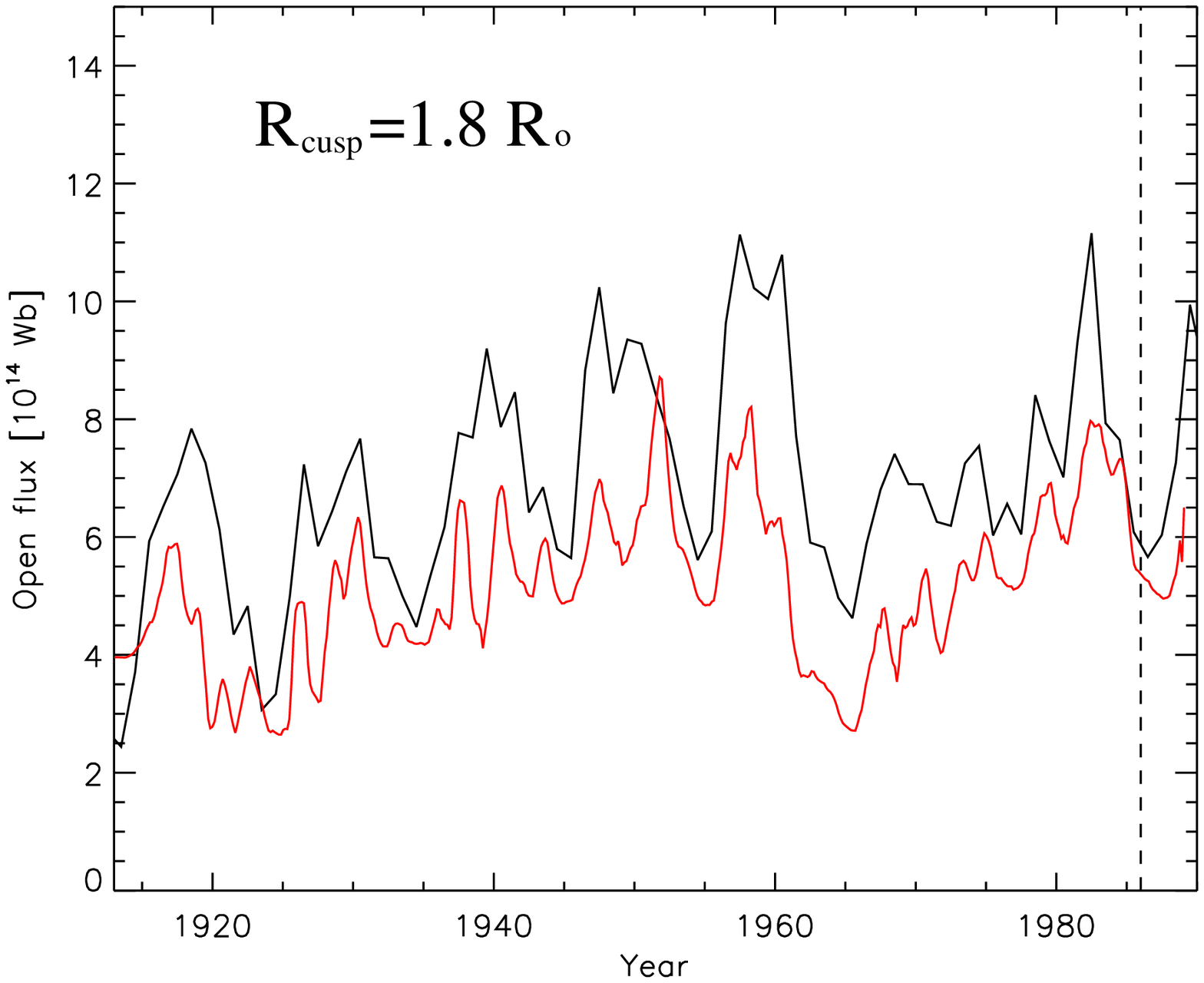}{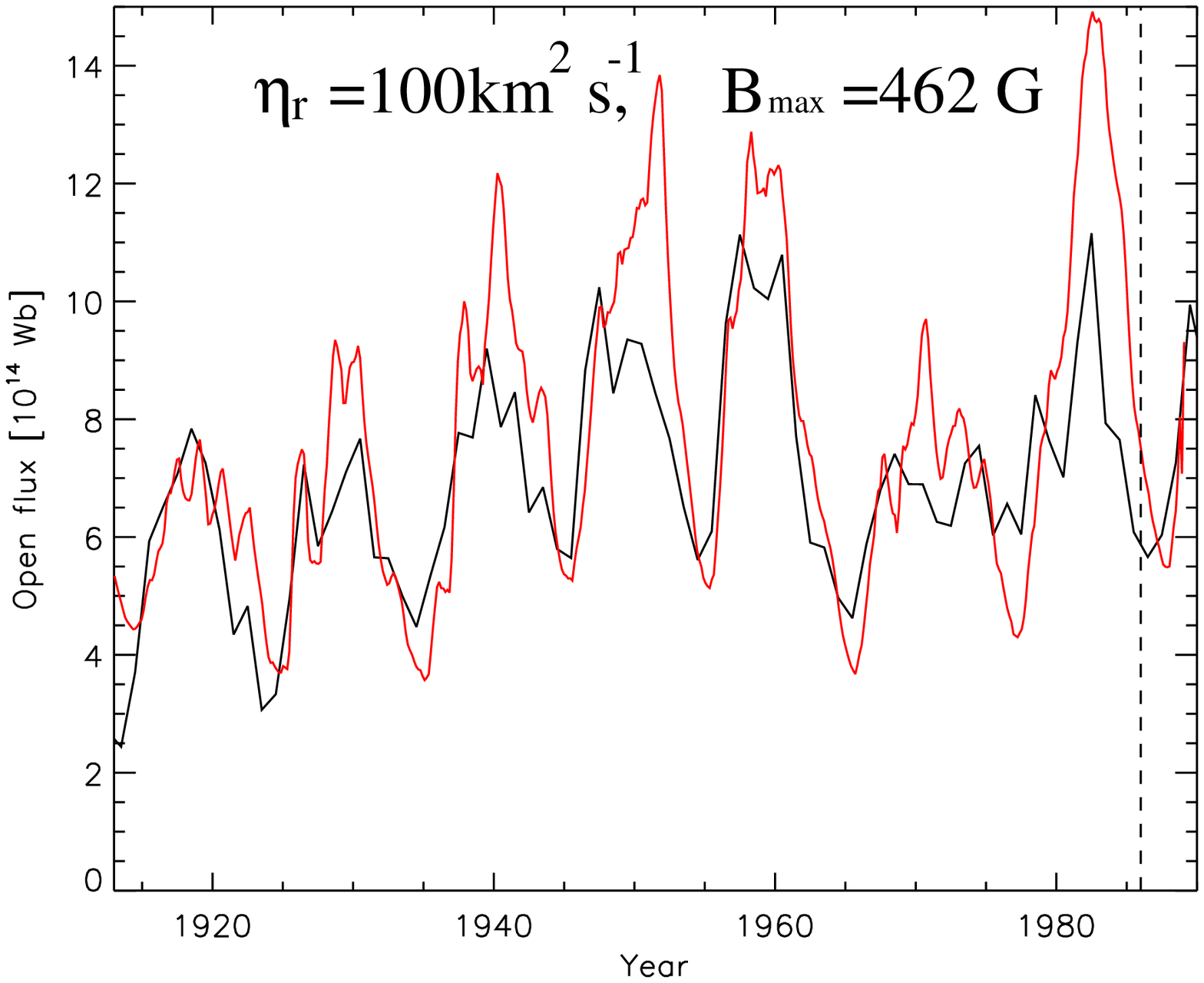}

\caption{Effect on \of~ of varying the parameters of the flux transport model. The black curve in each case is the
observationally inferred open flux while the red curve shows the
value from the simulation. For all other parameters the reference values were used.
The parameters varied were:\newline
\mbox{}\hspace{1cm} upper left: initial field strength $B_0$ ($-10.2$~G $\rightarrow$ $-8.5$~G),\newline
\mbox{}\hspace{2cm} right: surface diffusivity $\eta_H$ (250~km$^2$s$^{-1}$ $\rightarrow$ $500$~km$^2$s$^{-1}$),\newline
\mbox{}\hspace{1cm} middle left: radial diffusivity $\eta_r$ (0 $\rightarrow 100$ km$^2$s$^{-1}$ ),\newline 
\mbox{}\hspace{2cm} right: tilt angle reduction due to inflows $g$ (0.7$\rightarrow$1)\newline
\mbox{}\hspace{1cm} lower left: cusp surface radius $R_{\mathrm{cusp}}$ ($1.55 \rsun \rightarrow 1.8 \rsun$) \newline
\mbox{}\hspace{2cm} right: radial diffusivity $\eta_r$ ($0 \rightarrow  100$ km$^2$s$^{-1}$) and 
                       $B_{\mathrm{max}}$($374$ G $\rightarrow  462$ G) 
}
\label{fig_param}
\end{figure}

In the previous section we showed that a good representation of the empirically determined 
open flux \of~ could be achieved with the parameter values of the reference model.
Figure~\ref{fig_param} shows the effect of separately varying the
initial field strength, $B_0$, the surface diffusivity, $\eta_{H}$, radial diffusivity, 
$\eta_{r}$, and cusp surface height, $R_{\mathrm{cusp}}$. In all panels except the lower right we have
kept $B_{\mathrm{max}}$ constant: in this panel we recalibrated $B_{\mathrm{max}}$ to
account for the change in the total unsigned flux resulting from the change in $\eta_r$.

The upper left panel of Figure~\ref{fig_param} shows the effect of changing $B_0$. 
Since $B_0$ describes the initial polar field, varying this value leads to an offset in the strength 
of the axial dipole moment, which persists throughout the simulation when $\eta_r=0$. 
This offset results in alternating cycles having either stronger or weaker axial dipole moments 
depending on whether or not they have the same sign of dipole moment as that of the initial state. 
Near activity minima it is the lower order axial moments which dominate \of~ 
so that the minima alternately become higher and lower.

The upper right panel of Figure~\ref{fig_param} shows the effect of increasing $\eta_{H}$: 
the minima of \of~ are shifted upwards, while the maxima are not substantially affected.
The explanation for the upward shift is that $\eta_{H}$ determines the
amount of flux which crosses the equator and thus directly influences the axial dipole moment. 
There is also a weak but noticeable 22-year component, with the minima of alternating 
cycles being weaker. This 22-year component is present because we have not recalibrated 
$B_0$.
 
The  middle left panel of  Figure~\ref{fig_param} shows the effect of increasing $\eta_r$.
The enhanced decay of the field  not only
reduces \of~ around the minima but also during the rise phase a
cycle. This leads to too low minima and a delay of the  rising phase. There is 
again a 22-year component because $B_0$ has not been recalibrated.

The middle right panel shows the effect of varying the tilt angle reduction factor, $g$, 
from $0.7$ to 1, 
which modifies the magnitude of the polar fields and axial dipole 
moment. The signature is therefore an increase in the magnitude of the
changes in the dipole moment (and thus \of)  its minima, 
so that the effect almost cancels after two cycles.
This also produces a strong 22 year periodicity in the minima. 
We comment that $g=0.7$ is required to obtain the correct 
ratio between the maxima and minima of the open flux, as it essentially scales the  
low-order axial multipoles whilst barely affecting the equatorial multipoles.
Introducing $g$	does not affect	whether	or not the polar fields	reverse -- the 22 year 
periodicity, when $g$ is varied in isolation, can be removed by an appropriate choice of $B_0$.
 
The lower left panel shows that increasing $R_{\mathrm{cusp}}$ in isolation weakens \of. 
The effect is strongest during the maxima as it preferentially reduces the contribution
from higher order multipoles. The influence is thus qualitatively different from that of the other 
parameters in that it changes the relative contributions of the different multipoles. 

In the panels discussed so far we have kept $B_{\mathrm{max}}$, the scaling factor for the total flux of
newly emerging BMRs, constant. Varying $\eta_r$
as was done in the middle left panel changes the total amount of unsigned flux, and so 
affects the calibration of  $B_{\mathrm{max}}$. In the bottom right panel we therefore
show the effect of a change in $\eta_r$ together with the corresponding 
change in $B_{\mathrm{max}}$. Since the entire system is linear in $B_{\mathrm{max}}$,
changing  $B_{\mathrm{max}}$ merely rescales the result -- hence the result in the lower
right panel is just a scaled version of the result shown middle left panel. We note
that varying $\eta_{H}$ also affects the calibration.

This brief study of the 
effect of varying the parameters illustrates the kind of 
changes which occur. However, it does not rule out other choices
for the parameters which also could provide a good fit to the observations. In particular we
do not claim that non-zero values of $\eta_r$ are excluded, although we can say that,
at least for cycles 15--21, a good fit to the observations does not require a non-zero $\eta_r$.

\section{Conclusions}
The surface flux transport model including the effect of a cycle-dependent variation of the 
tilt angles of sunspot groups reproduces the major features of the observationally inferred open flux
 and the timing of the polar field reversals
from 1913 to 1986 (the period for which we have the tilt angle data).
The reversal of polar fields after strong cycles can be explained by the observed anti-correlation between
the active region tilt angle and cycle amplitude, so that no additional decay by radial diffusion was required to
achieve this result.

When the observed tilt angle are used, the polar field maxima from the model
are correlated with the strength of the following cycle. This correlation is 
likely to be present independent of the parameter choices,
provided the model reproduces the minima of the inferred open flux. The correlation 
suggests that the polar fields are an important ingredient of the solar dynamo process,
which is consistent with  Babcock-Leighton-type models.
The cycle-to-cycle variation of Joy's law might play a role in the 
nonlinear modulation of the solar dynamo.

\bibliographystyle{apj}
\bibliography{SFTC}


\end{document}